\documentclass[twocolumn,preprintnumbers,showpacs,amsmath,amssymb]{revtex4}
\usepackage{graphicx}
\usepackage{amssymb}

\begin{document}

\title{Device-independent quantum private comparison protocol without a
third party}

\author{Guang Ping He}
\email{hegp@mail.sysu.edu.cn}
\affiliation{School of Physics, Sun Yat-sen University, Guangzhou 510275, China}

\begin{abstract}
Since unconditionally secure quantum two-party computations are known to be
impossible, most existing quantum private comparison (QPC) protocols adopted
a third party. Recently, we proposed a QPC protocol which involves two
parties only, and showed that although it is not unconditionally secure, it
only leaks an extremely small amount of information to the other party. Here
we further propose the device-independent version of the protocol, so
that it can be more convenient and dependable in practical applications.
\end{abstract}


\pacs{03.67.Dd, 03.65.Ud, 03.67.Mn, 03.67.Ac, 03.65.Ta}

\maketitle


\section{Introduction}

Device-independent (DI) quantum cryptography has caught great interests
recently \cite{qi1065,qi481,qi996,qi997,qi998,qi992,di18}. It aims to
replace the model of the physical devices used in cryptographic protocols
with physically testable assumptions, e.g., the certification of
nonlocality. Thus the devices can be treated as black boxes that produce
outputs correlated with some inputs. This brings the advantage that the
assumptions needed to guarantee the security of the protocol can be
significantly reduced, so that the knowledge of the internal workings of the
devices is not required. The protocol remains reliable even if the devices
are provided by the adversary. Such a higher degree of security makes DI
protocols more dependable in practical applications than traditional quantum
cryptography.

Currently, most existing DI protocols focused on quantum key
distribution (QKD) \cite%
{qi1065,qi481,qi996,qi997,qi998,qi992,di18} and related tasks \cite{di247}, where all legitimate participants always collaborate
honestly against the attack of external cheaters. On the other hand, as
pointed out in Ref. \cite{di161}, DI two-party cryptography remains a
largely unexplored territory, as very few researches were dedicated to
multi-party secure computation problems, where the legitimate participants
do not trust each other since some of them may cheat. To our best knowledge,
Refs. \cite{di161,qbc29,qbc131,qbc110,qbc123,di205,di179} are the only
references on this field so far, which studied quantum bit commitment,
quantum coin tossing, weak string erasure, and position verification.

In this paper, we will take the first step towards the DI solution of
another multi-party secure computation problem -- quantum private comparison
(QPC), a.k.a. the socialist millionaire problem \cite{m2}. The goal of QPC
is to compare two secret numbers $a$ and $b$ of two parties Alice and Bob,
respectively, so that they finally learn whether $a$ and $b$ are equal,
without revealing any extra information on their values (other than what can
be inferred from the comparison result) to the other party. As a typical
example of multi-party secure computations, QPC plays essential roles in
cryptography, with many applications in e-commerce and data mining.

As it is well-known, unconditionally secure quantum two-party secure
computations are impossible \cite{qi149,qi500,qi499,qi677,qi725,qbc14},
i.e., the dishonest party can always obtain a non-trivial amount of
information on the secret data of the other party. QPC is also covered. To
circumvent the problem, almost all existing QPC protocols (\cite%
{qi1050,qi1043,qi1049,qi1120,qi1044,qi1045,qi1047,qi1116,qi1046,qi1048,qi1115,qi1110,qi1117,HeIJQI13,qi1192,qi1207,qi1250}
and the references therein) added a third party to help Alice and Bob
accomplish the comparison. The protocol in Ref. \cite{qi1203} is a rare
exception, in which such a third party is absent. But it was later found to
be completely insecure \cite{HeComment15,qi1354} because the entire secret
data of the honest party will always be exposed to the other party.

Very recently, we proposed a device-dependent (DD) QPC protocol which
involves two parties only \cite{HeDDQPC}, and showed that the average amount
of information leaked to the dishonest party can be as low as $14$ bits for
any length of the bit-string being compared. Therefore, despite that the
protocol is not unconditionally secure, the performance is quite good, and
this is achieved without relying on the third party. Here we will further
upgrade the protocol to the device-independent (DI) version, so that it can
enjoy even better security and versatility in practical applications.

The paper is organized as follows. In the next section, our previous DD
protocol will be briefly reviewed. Then in section 3, it will be turned into
the DI version, with its security analyzed in section 4.

\section{The DD protocol}

Let $H(x)$\ be a classical hash function which is a $1$-to-$1$\ mapping
between the $n$-bit strings $x$ and $y=H(x)$ (i.e., $H:\{0,1\}^{n}%
\rightarrow \{0,1\}^{n}$). Denote the two orthogonal states of a qubit as $%
\left\vert 0\right\rangle _{0}$\ and $\left\vert 1\right\rangle _{0}$,
respectively, and define $\left\vert 0\right\rangle _{1}\equiv (\left\vert
0\right\rangle _{0}+\left\vert 1\right\rangle _{0})/\sqrt{2}$, $\left\vert
1\right\rangle _{1}\equiv (\left\vert 0\right\rangle _{0}-\left\vert
1\right\rangle _{0})/\sqrt{2}$. That is, the subscript $\sigma =0,1$ in $%
\left\vert \gamma \right\rangle _{\sigma }$\ stands for two incompatible
measurement bases, while\ $\gamma =0,1$\ distinguishes the two states in the
same basis. In Ref. \cite{HeDDQPC}, the following protocol was proposed.

\bigskip

\textit{The DD QPC Protocol} (for comparing Alice's $n$-bit string $a\equiv
a_{1}a_{2}...a_{n}$\ and Bob's $n$-bit string $b\equiv b_{1}b_{2}...b_{n}$):

(1) Using the hash function $H(x)$, Alice calculates the $n$-bit string $%
h^{A}\equiv h_{1}^{A}h_{2}^{A}...h_{n}^{A}=H(a)$, and Bob calculates the $n$%
-bit string $h^{B}\equiv h_{1}^{B}h_{2}^{B}...h_{n}^{B}=H(b)$.

(2) From $i=1$ to $n$, Alice and Bob compare $h^{A}$ and $h^{B}$ bit-by-bit
as follows.

\qquad If $i$ is odd, then:

\qquad \qquad (2.1A) Alice randomly picks a bit $\gamma _{i}^{A}\in \{0,1\}$%
\ and sends Bob a qubit in the state $\left\vert \gamma
_{i}^{A}\right\rangle _{h_{i}^{A}}$.

\qquad \qquad (2.2A) Bob measures it in the $h_{i}^{B}$\ basis and obtains
the result $\left\vert \gamma _{i}^{B}\right\rangle _{h_{i}^{B}}$. He
announces $\gamma _{i}^{B}$\ while keeping $h_{i}^{B}$\ secret.

\qquad \qquad (2.3A) Alice announces $\gamma _{i}^{A}$.

\qquad If $i$ is even, then:

\qquad \qquad (2.1B) Bob randomly picks a bit $\gamma _{i}^{B}\in \{0,1\}$\
and sends Alice a qubit in the state $\left\vert \gamma
_{i}^{B}\right\rangle _{h_{i}^{B}}$.

\qquad \qquad (2.2B) Alice measures it in the $h_{i}^{A}$\ basis and obtains
the result $\left\vert \gamma _{i}^{A}\right\rangle _{h_{i}^{A}}$. She
announces $\gamma _{i}^{A}$\ while keeping $h_{i}^{A}$\ secret.

\qquad \qquad (2.3B) Bob announces $\gamma _{i}^{B}$.

\qquad (2.4) If $\gamma _{i}^{A}\neq \gamma _{i}^{B}$, then they conclude
that $a\neq b$, and abort the protocol immediately without comparing the
rest bits of $h^{A}$ and $h^{B}$. Otherwise they continue with the next $i$.

(3) If Alice and Bob find $\gamma _{i}^{A}=\gamma _{i}^{B}$ for all $%
i=1,...,n$ then they conclude that $a=b$.

\bigskip

Note that in the protocol, we compare the hash functions $h^{A}$ and $h^{B}$
instead of the secret strings $a$ and $b$ themselves. The purpose is to
change the information leaked to the other party from direct information
into mutual information on $a$ and $b$. It will not change the total amount
of information leaked.

\section{The DI protocol}

In each round of step (2) of the above protocol, $h_{i}^{A}$ and $h_{i}^{B}$
are compared in a non-entangled way. That is, one party (e.g., Alice)
prepares a qubit in the $h_{i}^{A}$ basis, then the other party (e.g., Bob)
measures it in the $h_{i}^{B}$ basis. This can be replaced with the
following entangled method. Alice prepares the Bell state $\left\vert \Phi
^{+}\right\rangle =(\left\vert 0\right\rangle _{0}\left\vert 0\right\rangle
_{0}+\left\vert 1\right\rangle _{0}\left\vert 1\right\rangle _{0})/\sqrt{2}%
=(\left\vert 0\right\rangle _{1}\left\vert 0\right\rangle _{1}+\left\vert
1\right\rangle _{1}\left\vert 1\right\rangle _{1})/\sqrt{2}$. She keeps the
first qubit and sends Bob the second one. Then they measure their qubits in
the $h_{i}^{A}$ or $h_{i}^{B}$ basis, respectively. Obviously, when $%
h_{i}^{A}=h_{i}^{B}$, their measurement results will always be equal, while
if $h_{i}^{A}\neq h_{i}^{B}$, their results will be different with
probability $1/2$. Thus the resultant protocol is equivalent to our original
DD protocol.

In DI cryptography, Bell states can further be replaced by pairs of devices
called nonlocal boxes, which can be supplied by either Alice or Bob, or even
other untrusty parties. These devices can be treated as DI black boxes that
take inputs and produce outputs, without the need nor the possibility to
check how they work internally. Each of Alice's (Bob's) boxes has four
inputs $S_{A}^{(0)}$, $S_{A}^{(1)}$, $S_{A}^{(2)}$, and $S_{A}^{(3)}$ ($%
S_{B}^{(0)}$, $S_{B}^{(1)}$, $S_{B}^{(2)}$, and $S_{B}^{(3)}$). For each
input, Alice's (Bob's) box can product either of the two outputs $\gamma
_{A}^{(k)}=\pm 1$ ($\gamma _{B}^{(k)}=\pm 1$), $k=0,1,2,3$. When the boxes
are manufactured honestly, they should act exactly like $\left\vert \Phi
^{+}\right\rangle $. For example, if we treat $\left\vert 0\right\rangle
_{0} $ and $\left\vert 1\right\rangle _{0}$ ($\left\vert 0\right\rangle _{1}$
and $\left\vert 1\right\rangle _{1}$) as the eigenstates of the Pauli
operator $\sigma _{z}$ ($\sigma _{x}$), then an honest implementation of
such a DI box pair can be realized using the Bell state $\left\vert \Phi
^{+}\right\rangle $ itself, with the inputs of the box at Alice's or Bob's
side being implemented with the measurements on her or his qubit,
respectively, as \cite{qbc123}%
\begin{eqnarray}
S_{A}^{(0)} &=&S_{B}^{(0)}=\sigma _{z},  \nonumber \\
S_{A}^{(1)} &=&S_{B}^{(1)}=\sigma _{x},  \nonumber \\
S_{A}^{(2)} &=&S_{B}^{(2)}=(\sigma _{z}+\sigma _{x})/\sqrt{2},  \nonumber \\
S_{A}^{(3)} &=&S_{B}^{(3)}=(\sigma _{z}-\sigma _{x})/\sqrt{2}.
\end{eqnarray}%
Define the Clauser-Horne-Shimony-Holt (CHSH) polynomials \cite{qi996}%
\begin{equation}
C_{1}=\left\langle \gamma _{A}^{(2)}\gamma _{B}^{(0)}\right\rangle
+\left\langle \gamma _{A}^{(2)}\gamma _{B}^{(1)}\right\rangle +\left\langle
\gamma _{A}^{(3)}\gamma _{B}^{(0)}\right\rangle -\left\langle \gamma
_{A}^{(3)}\gamma _{B}^{(1)}\right\rangle  \label{CHSH1}
\end{equation}%
and%
\begin{equation}
C_{2}=\left\langle \gamma _{A}^{(0)}\gamma _{B}^{(2)}\right\rangle
+\left\langle \gamma _{A}^{(1)}\gamma _{B}^{(2)}\right\rangle +\left\langle
\gamma _{A}^{(0)}\gamma _{B}^{(3)}\right\rangle -\left\langle \gamma
_{A}^{(1)}\gamma _{B}^{(3)}\right\rangle ,
\end{equation}%
where the correlator $\left\langle \gamma _{A}^{(k_{1})}\gamma
_{B}^{(k_{2})}\right\rangle $\ is defined as the probability $\Pr (\gamma
_{A}^{(k_{1})}=\gamma _{B}^{(k_{2})})-\Pr (\gamma _{A}^{(k_{1})}\neq \gamma
_{B}^{(k_{2})})$. If the boxes work as they were claimed, then the CHSH
values should reach the point of maximal quantum violation of the CHSH Bell
inequality, i.e., $C_{1}=C_{2}=2\sqrt{2}$. Also, when Alice and Bob choose
the same input $S^{(0)}$ (or $S^{(1)}$) for a pair of the boxes, they should
always obtain maximally correlated outputs $\gamma _{A}^{(0)}=\gamma
_{B}^{(0)}$ (or $\gamma _{A}^{(1)}=\gamma _{B}^{(1)}$). On the contrary, for
local boxes which have fixed output values without displaying any nonlocal
correlation, the CHSH values will satisfy the CHSH Bell inequality $%
C_{1}\leq 2$ and $C_{2}\leq 2$.

Also, like other DI cryptographic protocols (e.g., \cite{qi481,qi996,qi998}%
), it is assumed that Alice's and Bob's locations are secure, in the sense
that no unwanted information can leak out to the outside. Thus one party
cannot know the other's inputs and outputs to the DI boxes, unless the
latter party announces them.

With such nonlocal boxes, our protocol can be turned into the following DI
version.

\bigskip

\textit{The DI QPC Protocol} (for comparing Alice's $n$-bit string $a\equiv
a_{1}a_{2}...a_{n}$\ and Bob's $n$-bit string $b\equiv b_{1}b_{2}...b_{n}$):

(A) The check mode: Alice and Bob share many pairs of DI boxes and check
them as follows.

\qquad (i) Alice randomly chooses some of them, asking Bob to randomly pick
inputs into his boxes and announce both his inputs and outputs, then she
randomly picks inputs into her boxes and records the outputs.

\qquad (ii) Bob randomly chooses another portion of the box pairs, asking
Alice to randomly pick inputs into her boxes and announce both her inputs
and outputs, then he randomly picks inputs into his boxes and records the
outputs.

\qquad In both (i) and (ii), each of them check whether they have correlated
outputs when they picked the same inputs $S^{(0)}$ or $S^{(1)}$ for the
boxes of the same pair, and use the outputs they obtained when picking
different inputs to check whether the CHSH values $C_{1}$ and $C_{2}$
violate the CHSH Bell inequality.

(B) The compare mode: Alice and Bob randomly pick some of the rest unchecked
pairs of boxes to continue with the following steps.

\qquad (1) Using the hash function $H(x)$, Alice calculates the $n$-bit
string $h^{A}\equiv h_{1}^{A}h_{2}^{A}...h_{n}^{A}=H(a)$, and Bob calculates
the $n$-bit string $h^{B}\equiv h_{1}^{B}h_{2}^{B}...h_{n}^{B}=H(b)$.

\qquad (2) From $i=1$ to $n$, Alice and Bob compare $h^{A}$ and $h^{B}$
bit-by-bit as follows.

\qquad \qquad If $i$ is odd, then:

\qquad \qquad \qquad (2.1A) Alice randomly picks a DI box, inputs $%
S_{A}^{(h_{i}^{A})}$ and records the output $\gamma _{A}^{(h_{i}^{A})}$.
Then she tells Bob which one she picked.

\qquad \qquad \qquad (2.2A) Bob inputs $S_{B}^{(h_{i}^{B})}$ into his
corresponding box from the same pair and obtains the output $\gamma
_{B}^{(h_{i}^{B})}$. He announces $\gamma _{B}^{(h_{i}^{B})}$\ while keeping
$S_{B}^{(h_{i}^{B})}$\ secret.

\qquad \qquad \qquad (2.3A) Alice announces $\gamma _{A}^{(h_{i}^{A})}$.

\qquad \qquad If $i$ is even, then:

\qquad \qquad \qquad (2.1B) Bob randomly picks a DI box, inputs $%
S_{B}^{(h_{i}^{B})}$ and records the output $\gamma _{B}^{(h_{i}^{B})}$.
Then he tells Alice which one he picked.

\qquad \qquad \qquad (2.2B) Alice inputs $S_{A}^{(h_{i}^{A})}$ into her
corresponding box and obtains the output $\gamma _{A}^{(h_{i}^{A})}$. She
announces $\gamma _{A}^{(h_{i}^{A})}$\ while keeping $S_{A}^{(h_{i}^{A})}$\
secret.

\qquad \qquad \qquad (2.3B) Bob announces $\gamma _{B}^{(h_{i}^{B})}$.

\qquad \qquad (2.4) If $\gamma _{A}^{(h_{i}^{A})}\neq \gamma
_{B}^{(h_{i}^{B})}$, then they conclude that $a\neq b$ and abort the
protocol immediately without comparing the rest bits of $h^{A}$ and $h^{B}$.
Otherwise they continue with the next $i$.

\qquad (3) If Alice and Bob find $\gamma _{A}^{(h_{i}^{A})}=\gamma
_{B}^{(h_{i}^{B})}$ for all $i=1,...,n$ then they conclude that $a=b$.

\bigskip

For clarity, in the above description we wrote the check mode and the
compare mode separately. But for better security, the two modes should
actually be mixed together. That is, Alice and Bob choose some DI boxes to
run the check mode first, and they choose one of the unchecked DI box and
shift to the compare mode to compare one bit of $h^{A}$ and $h^{B}$. Then
they choose some other DI boxes and run the check mode again, followed by
another round of the compare mode to compare the next bit of $h^{A}$ and $%
h^{B}$, and so on. The times for shifting modes are decided randomly by both
parties in turns. Otherwise, if the compare mode is run only after the check
mode is completed, the provider of the DI boxes may cheat by building a
secret timer into each box, so that they all act honestly like the entangled
state $\left\vert \Phi ^{+}\right\rangle $\ during the check mode, then
switch to some kinds of cheating mode (e.g., giving fixed outputs regardless
the input values) automatically at the time when the compare mode is
expected to begin.

Also, it is important that in step (2.1A) (step (2.1B)), Alice (Bob) should
finish input to her (his) box and record the output before telling the other
party which box is picked. Otherwise, if she (he) announces which box is
picked first, and postpone the input/output process to step (2.3) after the
other party completed the input/output in step (2.2), then it could also
bring security problems. That is, if the other party is dishonest and he
provides the DI boxes, he may build a remote control into each box, so that
they usually act like real nonlocal boxes that can certainly pass the CHSH
inequality check. But when he knows which box is picked for the compare
mode, he engages the remote control to turn the box into the cheating mode
which gives a fixed output that known to himself beforehand, so that he can
cheat with the method that we will describe below in the paragraph before
Theorem 1.

\section{Security}

Since unconditionally secure QPC is impossible when only two parties are
involved, we does not attempt to make the information leaked in our protocol
arbitrarily close to $0$. Instead, our goal is merely to make it stay at a
low level. That is, the cheating we are going to deal with is how the
dishonest Alice/Bob tries to increase his/her information on the other's
secret data.

There is surely no security problem to consider when $a=b$, because both
parties naturally know the secret data of each other from the comparison
result. Now let us study the case when the protocol outputs $a\neq b$.

An important feature of our protocol is that, if it aborts after running $m$
($1\leq m\leq n$) rounds of step (2), the last $n-m$ bits of $h^{A}$ and $%
h^{B}$ will not be compared any more. They will not be input into the DI
boxes, or enter the protocol in any other form at all. Thus it is obvious
that these bits remain completely secret to the other party. As a
consequence, the amount of mutual information leaked to each party is $m$
bits at the most.

Therefore, the goal of a dishonest party is to increase $m$. To do so,
he/she has to make the protocol abort as late as possible. Without loss of
generality, let us assume that Alice cheats, and suppose that she prepares
and supplies all the DI boxes, which is a case that benefits her the most so
that we can obtain a general upper bound of the successful cheating
probability.

It is easy for Alice to cheat in each of the odd rounds, because in step
(2.3A) she can always announce $\gamma _{A}^{(h_{i}^{A})}=\gamma
_{B}^{(h_{i}^{B})}$\ even if the actual result is $\gamma
_{A}^{(h_{i}^{A})}\neq \gamma _{B}^{(h_{i}^{B})}$. This ensures that the
protocol will never abort at these rounds.
But in each of the rest $k\equiv m/2$ even rounds among the first $m$ rounds
of step (2), she is required to announce $\gamma _{A}^{(h_{i}^{A})}$\ in
step (2.2B) before Bob announces $\gamma _{B}^{(h_{i}^{B})}$ in step (2.3B).
Since she wishes to announce a value that satisfies $\gamma
_{A}^{(h_{i}^{A})}=\gamma _{B}^{(h_{i}^{B})}$\ so that the protocol will not
abort, she needs to guess the $\gamma _{B}^{(h_{i}^{B})}$\ value that Bob
will obtain from his DI box. Now let us prove that the probability $%
p_{guess} $ for her to make a correct guess cannot equal exactly to $1$.

\bigskip

\textbf{Theorem 1.} If all of Bob's DI boxes always give the same output $%
\gamma _{B}^{(0)}=\gamma _{B}^{(1)}$ no matter he inputs $S_{B}^{(0)}$\ or $%
S_{B}^{(1)}$, then they cannot pass the CHSH inequality check.

\textbf{Proof.} Recall that step (ii) of the check mode requires Alice to
announce both her inputs and outputs to a DI box when Bob has not input
anything into his corresponding box yet. Therefore, for these boxes, Alice
cannot monitor Bob's announced values of his $\gamma _{B}^{(0)}$'s and $%
\gamma _{B}^{(1)}$'s first, then to\ fake her\ output values $\gamma
_{A}^{(2)}$'s and $\gamma _{A}^{(3)}$'s to make them pass the CHSH
inequality check. Consequently, if all boxes give $\gamma _{B}^{(0)}=\gamma
_{B}^{(1)}$, there will be $\left\langle \gamma _{A}^{(2)}\gamma
_{B}^{(0)}\right\rangle =\left\langle \gamma _{A}^{(2)}\gamma
_{B}^{(1)}\right\rangle $ and $\left\langle \gamma _{A}^{(3)}\gamma
_{B}^{(0)}\right\rangle =\left\langle \gamma _{A}^{(3)}\gamma
_{B}^{(1)}\right\rangle $. Substituting them into Eq. (\ref{CHSH1}), we
immediately obtain%
\begin{equation}
C_{1}=2\left\langle \gamma _{A}^{(2)}\gamma _{B}^{(0)}\right\rangle \leq 2,
\end{equation}%
which is far below the correct expected value $2\sqrt{2}$. Consequently,
when Bob checks the $C_{1}$\ value given by the boxes picked in step (ii),
he will catch Alice cheating. This ends the proof. 

\bigskip

Of course, dishonest Alice does not have to make all of Bob's DI boxes
always give the same output $\gamma _{B}^{(0)}=\gamma _{B}^{(1)}$. She can
mix a small number of such boxes with real nonlocal boxes (i.e., these act
exactly like $\left\vert \Phi ^{+}\right\rangle $). With this method, the
corresponding CHSH value $C_{1}$\ may merely deviate slightly from $2\sqrt{2}
$, so that the cheating could be covered by statistical fluctuation. But
then it is obvious that once Bob randomly picks a box in step (2.1B), this
box does not necessarily be one of these which give fixed output $\gamma
_{B}^{(0)}=\gamma _{B}^{(1)}$ regardless Bob's input. Once Bob picks a box
whose output $\gamma _{B}^{(h_{i}^{B})}$\ depends on his input $%
S_{B}^{(h_{i}^{B})}$, the no-signaling principle prevents Alice from knowing
$\gamma _{B}^{(h_{i}^{B})}$ with certainty before Bob announces it. Thus
Theorem 1 leads us to the following conclusion.

\bigskip

\textbf{Corollary.} In step (2.2B), the probability $p_{guess}$ for
dishonest Alice to make a correct guess on the output value $\gamma
_{B}^{(h_{i}^{B})}$ that Bob will announce in step (2.3B) cannot equal to $1$%
.

\bigskip

In this case, the upper bound of the average amount of information on Bob's
secret data $b$ that leaked to Alice is calculated as follows. Continue with
the analysis before Theorem 1. As Alice has probability $p_{guess}$ to make
a correct guess on Bob's $\gamma _{B}^{(h_{i}^{B})}$, she can announce $%
\gamma _{A}^{(h_{i}^{A})}=\gamma _{B}^{(h_{i}^{B})}$ in step (2.2B),\ so
that in each of the first $k\ $even rounds there is probability $p_{guess}$
that the protocol will not abort. Consequently, the probability for the
protocol to abort at the $m$th round (i.e., it happens to continue for the
first $k-1$ even rounds while aborts at the $k$-th even round) is%
\begin{equation}
p_{abort}^{m}=p_{guess}^{k-1}(1-p_{guess}).  \label{p abort}
\end{equation}%
That is, with probability $p_{abort}^{m}$\ the protocol will abort at the $m$%
th round, and dishonest Alice will know nothing about the rest $n-m$ bits of
Bob's hash value, so that she learns $m$ bits of information at the most.
Also, with probability $p_{guess}^{[n/2]}$ the protocol does not aborts
until all bits are compared (here $[n/2]$ means the integer part of $n/2$).
In this case she learns all the $n$ bits. Summing over all possible $m$
values and recall that $m=2k$ (as the protocol will not abort at the odd
rounds when Alice cheats), the average amount of mutual information leaked
to dishonest Alice is bounded by%
\begin{eqnarray}
I_{A} &=&\sum_{k=1}^{[n/2]}m\times p_{abort}^{m}+n\times p_{guess}^{[n/2]}
\nonumber \\
&=&\sum_{k=1}^{[n/2]}2k\times p_{guess}^{k-1}(1-p_{guess})+n\times
p_{guess}^{[n/2]}.  \label{IA}
\end{eqnarray}%
Note that this is merely a upper bound, as we have not considered the cases
where dishonest Alice cannot even escape the detection in the check mode.
That is, Alice may reach this bound only if her DI boxes can pass the check
mode with the maximum probability $1$.

When taking into consideration the probability for Alice to pass the check
mode, the maximum value of $p_{guess}$ will also be bounded. But calculating
the exact bound could be complicated, because the CHSH values are merely
statistical results. In practice we cannot expect to find $C_{1}=C_{2}=2%
\sqrt{2}$\ exactly. Some statistical fluctuation has to be allowed. Noise
and manufacture imperfections of the experimental devices could also
affected the actual outcome of the CHSH values. Thus it is hard to obtain a
general result without knowing the specific performance of the experimental
devices used for the implementation of the protocol. Here, take for example,
let us follow Ref. \cite{qi992} to take $C_{1},C_{2}\geq 2.5$ as acceptable
values, and neglect the noise and device imperfections. In this case,
suppose that Alice prepares $61\%$\ of the DI boxes as real nonlocal boxes
(which have the theoretical expected values $C_{1}=C_{2}=2\sqrt{2}$), while
the rest $39\%$ boxes are prepared as local boxes which will produce fixed
output values known beforehand to herself regardless Bob's input (which have
$C_{1}=C_{2}=2$). Then the theoretical expected average CHSH values will be $%
61\%\times 2\sqrt{2}+39\%\times 2\simeq 2.505$, so that she stands a
nontrivial probability to pass the check mode. The local boxes enable Alice
to guess Bob's output $\gamma _{B}^{(h_{i}^{B})}$ with probability $1$. But
for the real nonlocal boxes, as shown in Eq. (5) of Ref. \cite{HeDDQPC}, the
maximum probability for her to guess Bob's output is $p_{\max }=\cos
^{2}(\pi /8)\simeq 0.8536$. In this case we have%
\begin{equation}
p_{guess}=61\%\times 0.8536+39\%\times 1\simeq 0.91.
\end{equation}

In Fig.1 we take $p_{guess}=0.91$ and show the probability for the protocol
to abort at exactly the $m$th round (i.e., $p_{abort}^{m}$ in Eq. (\ref{p
abort})) when Alice cheats. We can see that most of the time the protocol
will abort very soon. Thus the corresponding amount of information leaked
will be small. The cases that the protocol can last many rounds will occur
with extremely small probabilities only. Therefore, there is very low
chances that dishonest Alice can gain a large amount of information on Bob's
secret data.


\begin{figure}[tbp]
\includegraphics{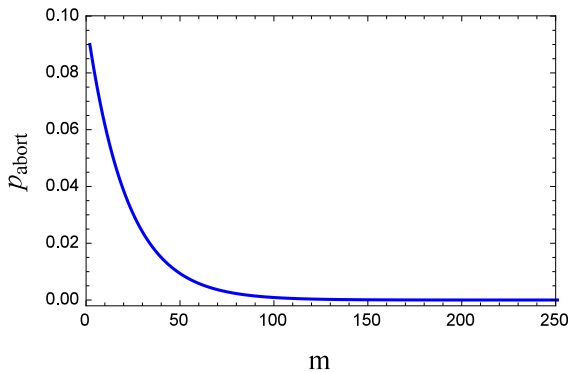}
\caption{The probability $p_{abort}^{m}$ for the protocol to abort at
the $m$th round when Alice cheats and $p_{guess}=0.91$.}
\label{fig:epsart}
\end{figure}


Fig. 2 shows the average amount of information leaked to Alice (i.e., $I_{A}$%
\ in Eq. (\ref{IA})) as a function of the length $n$ of the bit-strings $a$
and $b$ being compared. We can see that $I_{A}$\ never excesses $23$ bits
for any length $n$ of $a$ and $b$ when $p_{guess}=0.91$. It is a little
higher than that of the DD version, whose upper bound of the amount of
information leaked to Alice is $14$ bits, as proven in Ref. \cite{HeDDQPC}.
This is because in the DI protocol, $39\%$ of the DI boxes can be local
boxes which give fixed outputs, so that dishonest Alice's $p_{guess}$ can go
beyond the $p_{\max }$ in the DD version. But we should note that both $23$
bits (for the DI protocol) and $14$ bits (for the DD protocol) are merely
loose upper bounds. This is because, as elaborated in Ref. \cite{HeDDQPC},
when dishonest Alice saturates the maximum probability $p_{\max }$ for
guessing Bob's output $\gamma _{B}^{(h_{i}^{B})}$ in the DD protocol, there
is no known method for her to learn Bob's input $S_{B}^{(h_{i}^{B})}$\ with
probability $1$. Thus she cannot gain exactly $m$ bits of information on $%
h^{B}$ (and therefore Bob's string $b$) when the protocol aborts in the $m$%
th round of step (2). The same thing is also true for the local boxes. That
is, while they can enable Alice to guess Bob's output $\gamma
_{B}^{(h_{i}^{B})}$\ with probability $1$, they cannot provide her the
information on Bob's input $S_{B}^{(h_{i}^{B})}$ any more. It is worth
studying what will be the tight upper bounds of the information leaked in
both the DD and DI protocols.


\begin{figure}[tbp]
\includegraphics{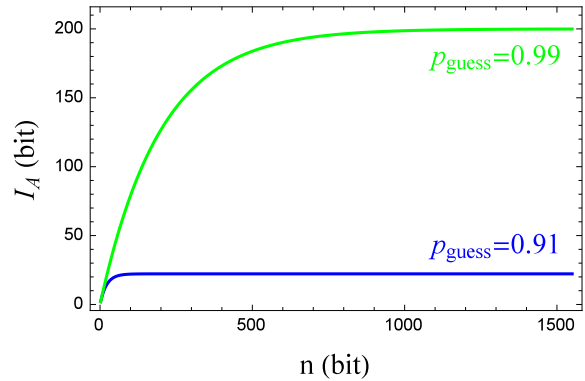}
\caption{The average amount of information leaked $I_{A}$ to
dishonest Alice as a function of the length $n$ of the bit-strings that they
compare.}
\label{fig:epsart}
\end{figure}





For illustration purposes, in Fig. 2 we also plotted $I_{A}$\ as a function
of $n$ when $p_{guess}=0.99$. It shows that $I_{A}$\ never excesses $200$
bits for any length $n$ even for such a high $p_{guess}$\ value. Considering
that most data we use in real applications nowadays are generally at the
size of megabytes to gigabytes, leaking only $200$ bits is not serious at
all. We would also like to emphasize that there is no known method to create
DI boxes which can pass the CHSH test with a nontrivial probability while
reaching $p_{guess}=0.99$ simultaneously. But even if such \textquotedblleft
black magic\textquotedblright\ exists, our result shows that the security
level of the DI QPC protocol is still quite acceptable.

From the symmetry of the protocol, it is trivial to show that the same
conclusion also applies if Bob (instead of Alice) cheats.

\section{Summary}

Thus we showed that although our DI QPC protocol is not unconditionally
secure, the average of the amount of information leaked to the dishonest
party is very low, and it has a fixed upper bound for any length of the
secret strings being compared. Also, the elimination of the third party
surely enhances the convenience for the practical applications of QPC.

This result also serves as yet another example that two-party cryptography
can enjoy the advantage of the DI scenario too, so that its security can be
even more reliable than their DD counterparts. It is also interesting to
study in future works whether the method can be applied to quantum private
query (QPQ) \cite{qi599,qi1536,qi1537}, which is another kind of multi-party
secure computation protocols that became a hot topic recently due to its
great practical significance.







\end{document}